\documentclass[runningheads]{llncs}
\usepackage{amsmath,amssymb,amsfonts}
\usepackage{algorithmic}
\usepackage{graphicx}
\usepackage{textcomp}
\usepackage{xcolor}
\usepackage{todonotes}
\usepackage{soul}
\usepackage{tikz}
\usepackage{hyperref}
\usepackage[utf8]{inputenc}
\usetikzlibrary{calc}
\usepackage[
backend=biber,
sorting=none,
citestyle=numeric
]{biblatex}
\begin{document}

\title{Decision process for blockchain architectures based on requirements}

\authorrunning{Six}
\titlerunning{Decision process for blockchain architectures based on requirements}
\title{Decision process for blockchain architectures based on requirements}
\author{Nicolas Six\orcidID{0000-0001-7563-3628}}

\institute{Centre de Recherche en Informatique (CRI),\break
Université Paris 1 Panthéon-Sorbonne, France\break
\email{nicolas.six@univ-paris1.fr}
}
\maketitle

\begin{abstract}
\label{Context and motivation}
In recent years, blockchain has grown in popularity due to its singular attributes, enabling the development of new innovative decentralized applications. 
\label{Question/Problem}
But when companies consider leveraging blockchain for their applications, the plethora of possible choices and the difficulty of integrating blockchain into architectures can hinder its adoption.
\label{Principal ideas/results} 
Our research project aims to ease the adoption of blockchain into companies, notably with the construction of an automated decision process to solve this issue in which requirements are first-class citizens, a knowledge base containing architectural patterns and blockchains refined over time, and an architecture generator able to process outputs into architectural stubs.
\label{Contribution}
This paper will also present our current progression on this decision process, by introducing the preliminary version that is able to choose the most suitable blockchain between multiple choices and our process-driven benchmarking tool.
\end{abstract}


\section{Introduction}

Building software is quite straightforward in a small project, but software architects can struggle to cope with complexity induced by the number of system components. Fortunately, they can take profit from existing architectural patterns, that are general and reusable solutions for implementing software \cite{monroe1997architectural} to help them to design their software. 
In parallel, blockchain has been a rising topic in the scientific and industrial communities over the last few years. 
It consists of a distributed database of interconnected blocks, accessible only for reading and appending. 
Indeed, altering blockchain retrospectively is hard, as every block is linked to the previous one and their integrity is ensured by the nodes that store a copy of the blockchain. 
First blockchain implementations were based on cryptocurrencies, such as Bitcoin \cite{nakamoto2008peer}. In a second time, some blockchains were built to support smart-contracts, that are instances of code embedded within the blockchain that can perform operations and hold states \cite{szabo1997formalizing}. 

In the field of Information Systems (IS), architects have to choose among a wide range of architectures and components that meet all of their requirements \cite{kornyshovathesis}. This is facilitated by field-tested patterns available from academic literature or companies' feedbacks. Thus, they have to make many decisions: which architecture to use, what components to include, and how to configure them to function accordingly to the requirements.
But when it comes to blockchain adoption, no field-tested blockchain architectural pattern exists yet and its integration into architectures requires deep knowledge about the technology.

Multiple studies tried to address this issue by introducing blockchain decision models to help to choose whether adopting blockchain or not and if yes, which type. In \cite{koens2018blockchain}, authors listed existing decision models and schemes such as \cite{wust2018you}. In \cite{belotti2019vademecum}, authors have built a comprehensive guide for blockchain technology including characteristics and parameters, then introduced a decision model based on specific questions and schemes. Those studies are useful to get a first idea about adopting a specific type of blockchain or not, but they are not diving into details.
Some studies explored the field of multi-criteria decision making (MCDM) to make decisions between blockchains (\cite{Farshidi2020}, \cite{Tang2019}). However, this only addresses the choice between blockchains, and those tools can be hard to use for architects non-expert in blockchain. 
Therefore, choosing a blockchain and fitting it into system architectures is still difficult, especially when taking a lot of requirements and assets into account, since existing methods lack an automated way to process them.

Through a doctoral thesis, we want to address those problems by building an automated decision process that determines the most suited architectural pattern for a given case. To this end, one needs multiple types of inputs inserted into the decision process, such as requirements or business assets. A major part of this work will be to build a knowledge base, from academic studies, expert feedback, or other information sources, and to update it over time. We also propose an architecture generator, to ease the adoption of blockchain architectures by automating the creation of architectural stubs.
The paper will be organized as follows: the Section \ref{researchgoals} will introduce our research questions and goals to address mentioned problems and research gaps, Section \ref{methodology} will present our research methodology, Section \ref{approach} will introduce the current chosen approach, and Section \ref{preres} will show current doctoral thesis progress. 

\section{Research questions and goals\label{researchgoals}}

The problematic mentioned in the introduction can be translated into multiple research questions that need to be addressed to ease blockchain adoption:

\renewcommand{\labelitemi}{}
\begin{itemize}
    \item RQ1: How can we extract and structure the data of business processes, assets, and requirements?
    \item RQ2: How can we process those data to make decisions regarding system architecture?
    \item RQ3: What are the impacts of blockchain specificity in the expression of the architectural model and the domain?
    \item RQ4: How blockchain can be integrated into existing architectural patterns?
\end{itemize}

To address those questions, we articulate this doctoral thesis around the construction of an automated decision process named the BLADE project (for BLockchain Automated DEcision process) where requirements are first-class citizens. The construction of this decision process will not be easy, as it raises a tremendous amount of locks. The major ones could be to find a method to compare every decision process inputs (infrastructure, requirements, assets, etc.) with architectural patterns in the knowledge base to find the pattern that fits best those inputs, as those elements are different from each other. Validating the decision process results' accuracy will also be difficult.

To extend the decision process, another goal of this doctoral thesis is to use recommendations and previous inputs to create architectural stubs, including entities, automated setup files, and interfaces. This aims to give architects a clean pattern to develop their applications, since implementing blockchain into architectures is a difficult task, but also to reduce the time between development and deployment, in a logic of cost-saving.

\section{Research methodology\label{methodology}}

This doctoral work will be based on the Design Science Research (DSR) methodology \cite{Hevner2004}. DSR is an incremental process: from the needs of people and (business) organizations, research is used to build artifacts that aim to address those needs, using methodologies and foundations from a knowledge base. After the completion of this part, an evaluation is performed on artifacts through case studies, experiments, simulation, and/or field studies to ensure that they correctly address business needs. Evaluation of artifacts is also crucial as it brings knowledge that serves to improve the knowledge base and the artifacts.

As the goal of this Ph.D. thesis is being able to determine the most suitable blockchain and architectural pattern, we need to know in detail their characteristics. To do that, we will incrementally construct our knowledge base, using previous studies on architectural and blockchain topics (benchmark results, methods, performance simulations, architectural patterns details, ...) as a first step, and then by performing studies on those topics by ourselves. 
Each update will allow us to incrementally refine our artifacts to be more accurate and support more blockchains or architectural patterns.

\section{The automated decision process\label{approach}}

Building such a process is a complex task, as there is a tremendous amount of possible inputs, plus the domain of software architecture is constantly evolving with new patterns and practices.
To get an architectural recommendation fitting with most of the initial requirements, we must be able to submit as many company requirements as possible and other valuable inputs in our decision process, such as business processes, existing architectures, or infrastructures.
Moreover, as technologies evolve, we have to feed our decision process with new data to keep it accurate from available architectural patterns.
In this section, we cover possible inputs of such a process, we introduce the processing part of it and associated locks that will have to be lifted, and we present our methods to construct then refine the knowledge base of the process. Finally, we introduce an architecture generator based on the decision process result.

\subsection{Inputs and processing}

The decision process uses 3 types of inputs to compute the most suitable architectural pattern and blockchain for a given case. The first one is the requirements. This input is defining desired non-functional requirements (eg. throughput, latency, ...) and functional requirements of chosen architecture. They can be expressed as strict requirements ("must have") or preferences ("better if the system has that"). Preferences are rated on a Likert scale \cite{allen2007likert} from 0 (Indifferent), to 1 (Extremely desirable). 
The second input considered are business processes (for IS). As they represent the logic of software that will use blockchain and related architectural patterns, they will be used to evaluate the compatibility of decision process choices with them. For example, a possible evaluation would be to verify if performance requirements are met for a specific business process when using a blockchain architecture. Business processes could even embed requirements introduced before, as requirements could differ depending on chosen business processes.
Enterprise assets are the third possible input. Existing infrastructures (eg. servers, network equipment, cloud services, ...), human skills on specific technologies, or programming patterns expertise could refine the choice as reuse enterprise assets would cut on implementation costs.   

Inputs processing using the knowledge base is one of the biggest locks that will have to be lifted during this doctoral thesis. Indeed, as inputs are different from each other regarding their unit, scale, or type, it is difficult to
compare them. A possible solution to address this problem is the utilization of multi-criteria decision making methods \cite{kornyshova2007mcdm}. 
Also, blockchain and architectural patterns performances may vary depending on the infrastructure and concerned business processes. Determining those characteristics will require to perform automatic benchmarks, a difficult task if the company does not have a test infrastructure available for that or if they don't know exactly their infrastructure topology (such as IoT infrastructures). Simulations could also be considered, but that requires to find accurate formulas. 
In the end, the decision process will increment over the list of stored architectural patterns and return for each of them a fitting score from 0 to 1, determining which pattern is the most suitable to fulfill most of the requirements and match with company assets and business processes. 

\subsection{Knowledge base construction and continuous improvement\label{ci}}

Following our methodology and our plans for this doctoral thesis, one of the biggest steps will be the construction of a relevant knowledge base, where data comes from multiple sources.
First, academic studies will help us to build this knowledge base and the decision process (eg. case studies, state-of-the-art, benchmarks, ...). They will provide empirical data about blockchains attributes and performance, requirements elicitation and weighting, and architectural implementation examples.
In a second step, interviews with experts or software architects will be a major advantage to strengthen our decision process. We identified 2 types of potential interviews. The first one is the peer-review of the decision process where interviewees could deep dive into the functioning and make comments on it to fix process flaws right before going on further tests. The second one is solving a sample test case study using their experience and knowledge, then compare their choices to the decision process and architecture generator results. We would obtain a bias between interviewees' results and ours that could be discussed to understand why different choices were made. 

The decision process will have to be improved over time following either the creation of new architectural patterns or studies released after the first version.
In addition to that, we are expecting to challenge our decision process on real-life use-cases with experts.
We plan to ask companies that plan to build a new application to run the decision process and get in return architecture stubs, that they will use to build their software. Using monitoring tools, we would be able to collect execution traces and analyze them to identify potential overheads or architectural flaws then improve the decision process and the architecture generator. 
Finally, to keep the decision process accurate, we need a relevant up-to-date knowledge base. To do so, we envision collaborating with industrial and open source communities, that could benefit from our work. Also, inputs defined in the current knowledge base will be refined as a whole using companies' feedbacks but also by category depending on the use-case under consideration. 

\subsection{Architecture generation}

Getting architectural recommendations is the main goal of our project, but we plan to go one step further by adding a generator tool at the end of the decision process. 
With all the inputs that were submitted and the result of the decision process, we have all the information about all future layers of the architecture. 

For the application layer, one of the decision process inputs is a collection of BPMN files, that are modeling company business processes and thus applications that will have to run on the architecture. Thus, as it is possible to generate entities from them, we implemented a tool called BPMN Adapter available on Github\footnote{https://github.com/nherbaut/bpmn\-adapter}.
Regarding network and physical layers, as the infrastructure and the architectural pattern are known, it is possible to generate configuration and setup files for servers and network equipment, to define desired communication protocols, operating systems, or network interfaces.

This work is important as it aims to drastically reduce the time between software conception and production deployment by automating all possible steps and therefore cutting on costs.
This is also a way for us to get validations of our decision process, as one of our goals is having generated architectures implemented within real business case studies where execution traces would be monitored. This would indicate eventual bottlenecks and imperfections of generated architectures and thus help us improve our project artifacts. 

\section{Preliminary results\label{preres}}

\subsection{Automated decision process current progression}

Since the beginning of this work, progress has been made on the automated decision process and results in a study \cite{six:hal-02537280} that introduces its foundations and evaluates its results through a real-life case study. 
For the moment, the decision process is able to find the most suitable blockchain for a given set of non-functional requirements, expressed as presented in our approach (Section \ref{approach}). Selected requirements that can be entered as inputs have been selected as they can be grouped under the categories of ISO 25010 standard that defines macro-attributes to consider in a system or software to ensure its quality. The goal behind this choice was to ease the choice of blockchain technology by non-experts in this domain. 
After data has been entered into the system, the decision process eliminates every blockchain that does not match strict requirements, then evaluates the suitability of every blockchain contained in our knowledge base using TOPSIS. This is a multi-criteria decision-making method able to rank given alternatives by computing the most and the least desirable alternative, then evaluating the distance of each considered alternative between them \cite{lai1994topsis}. 

Our current knowledge base is constituted with the 4 most used blockchains (in businesses\footnote{https://www.hfsresearch.com/pointsofview/whos-winning-the-battle-of-enterprise-blockchain-platforms}) attributes, plus Bitcoin, currently considered as the most-known blockchain. Their characteristics have been extracted from multiple sources: studies, technical documentation, whitepapers, and other relevant sources.  
Current results are promising, as the application of our automated decision process to the case study mentioned before returned the same result as recommended by blockchain experts in the case study. 

\subsection{Business process driven benchmark framework}

As one of our goals is the improvement of our knowledge base with benchmark studies, we decided to implement a highly customizable benchmark framework based on business processes and workflows. 
With business processes (as XML files) or JSON workflow files as inputs, we are able to generate and deploy sample smart-contracts as microservices on blockchains to benchmark their performance on those environments. Sample (blockchain smart-contracts) microservices exposes 4 methods, that respectively try to solve a difficult dummy computing problem, an increment on a loop, store data, and read data. The "difficulty of completion" of each method can be chosen using a method parameter called difficulty. As it rises, the number of increments on the loop, the size read and stored and the difficulty of the computing problem increases. 

Our initial idea was to compare microservices architectures with blockchain technology (eg. Ethereum) by benchmarking them using our suite, but we broader the possibilities by implementing the support of custom smart-contracts benchmark. Instead of generating dummy smart-contracts, we can provide one. Therefore, we can benchmark blockchain performance over a dedicated smart-contract, by varying the number of nodes, their hardware, and the characteristics of the blockchain used (eg. block time, block size, consensus protocol, ...). This will help us to strengthen our knowledge base during this thesis. 

\section{Conclusion and future work}

The choice between multiples architectures is not trivial for enterprises.
A lot of inputs and features had to be taken into account when determining which architectural pattern will be chosen, an impossible task when non-automated. 
Moreover, the decision process could be biased with the software architect's personal experience.

In this paper, we are introducing our plans to build a decision process focused on blockchain architectures to address those issues.
From a set of inputs, that are requirements, business processes, and business assets, the decision process will be able to translate them into architectural suggestions, but also using these latter to generate architecture stubs (entities generation, communication protocols, interfaces, etc.). 
One of the main challenges of this project will be finding a way to translate all inputs and all architectural patterns, into a set of numeric values, but also weighting them in the decision process as all the features do not have the same importance. 
To do that, we proposed to use many available scientific-related studies, but also to ask software architects and experts for feedback to validate the decision process or refine it. 
We also introduced our plans to improve the decision process after the first release with an iterative approach, taking new studies as inputs and having it tested by companies on real-life projects. 
Finally, we present our preliminary results on this topic that is a decision process for blockchain technologies called the BLADE project. It is able to recommend the most suitable blockchain from non-functional software-related requirements and user preferences, but also our process-based benchmarking tool.

As we saw that in the decision process performances are expressed in wide intervals instead of fixed values, we plan in the near future to improve this aspect by finding a way to determine those values from the context (blockchain technologies and specific blockchain parameters) using our benchmarking tool to determine such formulas. Another planned work will be to study possibilities of integration of blockchain into existing architectural patterns notably by performing a literature review on blockchain and architecture, to be able to include them inside the decision process in the future. As we adopted a constructive research methodology, we plan to move forward following this step by step approach to construct the automated decision process that takes into account business assets and requirements. 

\section*{\ackname}
This doctoral thesis is supervised by Pr. Camille Salinesi and Dr. Nicolas Herbaut.

\printbibliography

\end{document}